\documentclass[compsoc, conference, a4paper, 10pt, times]{IEEEtran}

\usepackage{cite}
\usepackage{amsmath,amssymb,amsfonts}
\usepackage{algorithmic}
\usepackage{graphicx}
\usepackage{textcomp}
\usepackage{xcolor}
\usepackage{booktabs}
\usepackage[hidelinks]{hyperref}
\usepackage{paralist}
\usepackage{comment}
\usepackage{todonotes}
\usepackage{multirow}

\begin{document}

\title{Automating privacy decisions -- where to draw the line?}


\author{\IEEEauthorblockN{Victor Morel}
\IEEEauthorblockA{\textit{Chalmers University of Technology} \\
Gothenburg, Sweden \\
morelv@chalmers.se}
\and
\IEEEauthorblockN{Simone Fischer-Hübner}
\IEEEauthorblockA{\textit{Chalmers University of Technology} \\
\& Karlstad University \\
Gothenburg \& Karlstad, Sweden \\
simonefi@chalmers.se, simofihu@kau.se}
}

\maketitle

\begin{abstract}
Users are often overwhelmed by privacy decisions  to manage their personal data, which can happen on the web, in mobile, and in IoT environments.
These decisions can take various forms -- such as decisions for setting privacy permissions or privacy preferences, decisions responding to consent requests, or to intervene and ``reject'' processing of one's personal data --, and each can have different legal impacts.
In all cases and for all types of decisions, scholars and industry have been proposing tools to better automate the process of privacy decisions at different levels, in order to enhance usability.
We provide in this paper an overview of the main challenges raised by the automation of privacy decisions,
together with a classification scheme of the existing and envisioned work and proposals addressing automation of privacy decisions.
\end{abstract}




\section{Introduction}
\label{sec:intro}
Whilst facing a surge of data collection by various actors, data subjects are often overwhelmed by requests for privacy decisions alongside with the task to manage their own personal data.
\textbf{Privacy decisions} can take a variety of forms, each with its own implications.
Privacy decisions can consist of \textit{Privacy Permission Settings}, which are for instance used when managing mobile applications~\cite{google_permissions_nodate,apple_control_nodate}.
Other privacy decisions consist of \textit{Privacy Preference Settings}, which are indications of the users' privacy wishes, often used to support the creation of usable privacy notices as part of consent forms.
Users can also \textit{Consent} to the processing of their personal data,
in that case the privacy decision has a clear legal impact (Articles 4 (11), 6 (1) (a), 7 of the EU General Data Protection Regulation (GDPR)).
Lastly, users can \textit{Reject} the processing of their data under certain conditions, e.g. by withdrawing their consent (Article 7(3) GDPR), objecting to data processing (Article 21 GDPR) or opting out.

In all environments -- the web, the IoT, and mobile -- scholars and industry have been addressing the automation of these privacy decisions in order to facilitate their usable management. Especially in mobile and IoT environments involving devices with limited screen sizes or limited user interactions, the usability~\cite{iso_iso_2018} of privacy management is a key challenge that could be addressed with the help of automation.
This automation process can for instance take the form of \textit{cookie consent tools}~\cite{kladnik_i_nodate,nouwens_dark_2020,firefox_firefox_2022}
or \textit{privacy assistants}~\cite{das_personalized_2018},
which may leverage a range of techniques, from simple rules to state of the art machine-learning (ML).

Nevertheless, the automation of privacy decisions also raises ethical and legal questions, especially regarding autonomy and control of users over their data -- the latter being an essential privacy principle highlighted in Recital 7 GDPR.
These questions are of particular interest when decisions, such as consent according to Article 4 (2), require an active and affirmative behaviour from the user, which contradicts a fully automated approach. 
For example, we observe that certain cookie consent tools can consent on behalf of users without an explicit affirmative action~\cite{firefox_firefox_2022,kladnik_i_nodate}, and some proposals for privacy assistants suggest that consent could be fully automated based on observed privacy preferences of users~\cite{colnago_informing_2020}.

However, automation can also increase usable user control over the processing of personal data, e.g. by dynamically creating usable privacy notices as part of consent forms, which are ``concise, transparent, intelligible and easily accessible'' in line with Article 12 GDPR, or by enabling more fine-grained and contextual controls via ``dynamic consent''~\cite{schlehahn2020}.

Our objective is to foster an interdisciplinary debate and research for addressing this inherent 
tension of automation -- that could both limit and at the same time enhance user control --  with the aim of 
laying the foundations for finding and promoting
usable and legally compliant privacy decision tools utilising automation. 
More specifically, the research questions addressed by this paper are summarized as follows:
\begin{enumerate}
    \item What types and categories of privacy decision tools exist?
    \item Under which conditions is automation of privacy decisions (and of consent specifically) compliant with the GDPR (and other ethical principles)?
    \item To what extent can automated privacy decisions promote informed control in line with the GDPR and can benefit users?
\end{enumerate}

The focus of our work will be on privacy decisions of users for controlling the disclosure and processing of their personal data.

Drawing from legal principles related to decision making from EU data protection law (summarised in Section~\ref{sec:background}), the state of the art of technical solutions surveyed from scientific literature, reports and legal opinion papers.
This work provides a classification scheme of existent or possible technical automation solutions to help the reasoning about these approaches and their lawfulness (Section~\ref{sec:criteria}).
Finally, we conclude in Section~\ref{sec:discussions} with remarks including a discussion on the possible trade-offs and guidelines for using the scheme.
We posit that while automation can enhance usability of solutions assisting privacy decisions, fully automated decisions are almost always in conflict with legal requirements.
Therefore, only partially automated solutions appear to meet both usability and legal compliance.

In contrast to previous classifications (see e.g. \cite{colnago_informing_2020, Santos23}), this scheme initiates a richer set of categories of decisions and automation.
To the best of our knowledge, no related work attempted to provide an overview of the types and of the lawfulness of automation of various privacy decisions.
Papers addressing parts of these aspects, such as \cite{colnago_informing_2020, Santos23}, 
are referred to and/or discussed along the document.

\section{Legal background}
\label{sec:background}
As a basis for our classification and discussion, this section briefly summarises European legal rules of relevance for privacy decision making and automation.~\footnote{We focus specifically on EU jurisdiction because of the extra-territoral application of the GDPR -- the \textit{de facto} standard in personal data protection world-wide. However, note that some US laws -- such as CCPA/CRPA -- can be of interest in this context.}

\textbf{Legal requirements for consent.} Pursuant to the GDPR, personal data shall only be processed if at least one of the six legal grounds listed in Article 6 GDPR applies, such as consent of the data subject, among others. Consent needs to be \textit{informed}, \textit{specific}, \textit{freely given}, and \textit{unambiguous}, which entails a \textit{clear statement} or an \textit{affirmative action} (Article 4 (11) GDPR). 
The latter requirement also implies that silence or pre-ticked boxes do not lead to a valid consent (see Recital 32 GDPR). Neither does the absence of a reject button on the first layer (see the report of the European Data Protection Board (EDPB) Cookie Banner Taskforce~\cite{cookie_banner_taskforce_report_2023}). 
Note that the interpretation on what constitutes a valid consent is still a vivid debate, as also discussed in the updated guidelines on consent~\cite{wp29_guidelines_2020} by the EDPB.

\textbf{Revocable consent}.  Article 7(3) of the GDPR explicitly states that users must be able to withdraw consent at any time, and that it shall be as easy to withdraw as to give consent.

\textbf{Explicit consent.} \textit{Explicit} consent  is required for three cases that especially pose privacy risks:
\begin{itemize}
    \item when the personal data to be processed constitutes special categories of data, i.e., sensitive personal data (Article 9 (2) (a)),
    \item when personal data is processed for automated individual decision-making including profiling (Article 22 (2) (c)),
    \item for data transfers to third countries or international organisations in the absence of adequate
safeguards (Article 49 (1) (a)).
\end{itemize}

Whilst the GDPR is specific on the definition of special categories of data in its Article 9,
it is not directly defining conditions for an explicit consent. 
The European Data Protection Board (EDPB) however explains that whereas a ``regular consent'' already requires a ``statement of affirmative action'', an explicit consent additionally requires that the data subject gives an ``express statement of consent''~\cite{wp29_guidelines_2020}.

\textbf{Consent for tracking technologies.} The ePrivacy Directive (ePD)~\cite{ePD-09} complements the GDPR with more specific rules for electronic communication providers and requires that whenever cookies and other tracking technologies are stored and/or read from the user's device, the ePD (in Article 5(3)) requires controllers to request consent for the storage of such trackers for certain non-essential purposes for processing data (such as advertising).

\textbf{Right to object.}
The GDPR specifies in its Article 21 the right to object to the processing of  personal data in certain circumstances, including the right to object to direct marketing and profiling or to object in the cases where the legal ground for the processing is public interest or legitimate interest.
Of particular interest in our context is Article 21 (5) that states that this right may be exercised ``by automated means using technical specifications''.

\textbf{Data Protection by Design and by Default.} 
The GDPR specifies in its Article 25 that the controller must implement appropriate technical and organisational measures to safeguard privacy and data protection principles right from the start (\textit{by design}), and that personal data should be processed with the highest privacy protection (\textit{by default}).
These two obligations need to be considered when it comes to the automation of decisions, as it implies that user settings must by default automatically abide to the highest privacy protection.

\section{Classification scheme}
\label{sec:criteria}

For addressing our research questions, we propose a 2-dimensional scheme for grouping relevant existing or possible technical approaches for enabling privacy decisions with different degrees of automation.
This classification is not meant to be exhaustive, rather illustrative.
The two dimensions of our classification scheme are the following: 
1) the type of the privacy decisions (permissions/access control settings, privacy preference settings, consent, and reject) and 
2) their level of automation (manual, semi-automated, automated). 

As mentioned above, in the context of this work, we focus on privacy decisions of users for controlling the  disclosure and conditions for the processing of their personal data, which are thus directly or indirectly related to consent. These are decisions that users are frequently confronted with, and for which usability is a major challenge, as such privacy decisions usually only become the users' primary goal when they are exposed to them.

Other types of privacy decisions that users, on their own initiative, need to make for exercising their data subject rights pursuant to the GDPR are not considered within the scope of our work (except for the rights directly related to consent, such as the right to revoke consent and to object).

The different types of privacy decisions that we consider originate from a distinction between those with a direct legal implication -- consent and reject --, and those without -- permissions and privacy preferences --, the latter having been devised for technical systems and to improve transparency, respectively.
These four types of privacy decisions directly concern choices of users regarding access and use of their personal data by external entities, albeit to various degrees and with different implications.
In a nutshell: privacy preferences are used as an indication, permissions for purely technical settings, and consent and reject decisions entails legal decisions, as Section~\ref{sec:obj} will describe in more depth.
Both dimensions are explored in more detail below and summarized in Table~\ref{tab:summary}.

\begin{table*}[ht!]
\begin{tabular}{|c|l|l|l|}
\hline
\multicolumn{1}{|l|}{\begin{tabular}[c]{@{}l@{}}Level of automation $\rightarrow$\\ Type of decision $\downarrow$\end{tabular}} & \multicolumn{1}{c|}{Manual} & \multicolumn{1}{c|}{Semi-automated} & \multicolumn{1}{c|}{Fully automated} \\ \hline
Privacy permission & Mobile permissions (AOI)~\cite{google_permissions_nodate,apple_control_nodate} & \begin{tabular}[c]{@{}l@{}}Recommendations~\cite{bahirat_data-driven_2018,smullen_best_2020}\\ Mobile Privacy Assistant~\cite{liu_follow_2016}\\ PPL sticky policies~\cite{ardagna2009primelife}\\ Mobile permissions (AOFU)\end{tabular} & Dynamically granted permissions~\cite{wijesekera2017feasibility} \\ \hline
Privacy preferences & \begin{tabular}[c]{@{}l@{}}DNT~\cite{kamara_not_2016}\\ P3P~\cite{cranor_platform_2002}\\ PPL~\cite{ardagna2009primelife}\end{tabular} & \begin{tabular}[c]{@{}l@{}}On the fly policy management~\cite{angulo2012towards}\\ IoT Privacy Assistant~\cite{das_personalized_2018}\end{tabular} & \textit{May contradict DPbD principle} \\ \hline
User consent & \begin{tabular}[c]{@{}l@{}}\textit{Traditional consent forms}\\Data Custodian~\cite{edps_opinion_2016,european_union_agency_for_cybersecurity_data_2021,morel_enhancing_2020}
\end{tabular} & \begin{tabular}[c]{@{}l@{}}Negotiation~\cite{morel_enhancing_2020}\\ JITCTA~\cite{patrick2003privacy}     \\ Dynamic consent~\cite{schlehahn2020,asghar2012flexible}\end{tabular} & \begin{tabular}[c]{@{}l@{}}I don't care about cookies~\cite{kladnik_i_nodate}\\ Firefox Cookie Banner Handling~\cite{firefox_firefox_2022}\\ Data in escrow~\cite{colnago_informing_2020}\end{tabular} \\ \hline
Reject & \begin{tabular}[c]{@{}l@{}}GPC~\cite{noauthor_global_nodate}\\ Smart Places~\cite{future_of_privacy_forum_smart_nodate}\end{tabular} & \textit{In line with Article 21 (5) GDPR} & \begin{tabular}[c]{@{}l@{}}Consent-O-Matic~\cite{nouwens_dark_2020}\\ ADPC~\cite{human2021advanced}\\ TCF~\cite{iab_europe_tcf_nodate}\end{tabular} \\ \hline
\end{tabular}
\vspace{.2cm}
\caption{Summary table of illustrative examples used in the classification scheme.
The table is not meant to be exhaustive, but can provide an overview of various levels of automation applied to different types of privacy decisions.}
\label{tab:summary}
\end{table*}

\subsection{Type of decisions}
\label{sec:obj}
We consider four types of privacy decisions within the scope of our work. 
These types of decisions are not always clearly distinguishable from each other, and may therefore also partly overlap, as we also discuss below.

\subsubsection{Privacy permission settings}
Setting privacy permissions refers to settings of access control rules in systems for permitting access to one's data.
Such privacy permissions are typical of mobile phone operation systems, such as Android~\cite{google_permissions_nodate} or iOS~\cite{apple_control_nodate}.
The decision made by the user will determine  
the extent to which a controller or a processor can be granted access to certain personal data or not.
Note that privacy permissions on mobile operating systems usually require consent ``upon installation or during runtime''~\cite{european_union_agency_for_network_and_information_security_privacy_2017}, but it is not always necessarily the case. For instance, the legal basis of contract (Article 6 (1)(b)) can also apply, e.g. for a banking app to forward account information when transferring money~\cite{Art29WP13}.

\subsubsection{Privacy preference settings}
\label{sec:pref}
In contrast to permissions, privacy preferences are also often (and within the scope of this paper) defined as mere \textit{indications} of the privacy \emph{choices} made by the user that are not necessarily binding the controller, and may thus not be technically enforced at services side.
Privacy preferences -- e.g., as used for P3P~\cite{cranor_platform_2002}, the PrimeLife Policy Language PPL~\cite{ardagna2009primelife} and A-PPL \cite{azraoui2015ppl}, or Pilot~\cite{pardo_analysis_2019} more specific to the IoT -- 
are typically written in a machine-readable form allowing a policy engine to determine whether they match with a machine-readable privacy policy of the data controller. 
The extent to which a data controller's  policy matches the declared user's privacy preferences can be prominently be displayed, thereby contributing to more usable and transparent
privacy notifications. For instance the P3P privacy bird tool~\cite{cylab_usable_privacy_and_security_laboratory_privacy_nodate} used  colors (green, yellow, red), form and sound of a bird icon in the browser title bar to illustrate whether there is a match of the user's preferences with the side's policy, if there is no match, or if the site has no policy.

Privacy policy languages, such as P3P or PPL, that have been researched and developed 15-20 years ago, have not successfully been deployed in practice -- also due to the reason that they require support from the services sides that need to host machine-readable policies. Still, they provide important concepts for classifying privacy decisions and automation, as we will see below. 
Moreover, machine-readable and enforceable privacy policy languages still subject of research in recent EU projects, such as the TRAPEZE EU projects~\cite{bonatti2021representing}.

\subsubsection{Consent}
As discussed in Section \ref{sec:background}, 
consent is one of the legal grounds for making personal data processing legitimate pursuant to Article 6 GDPR.
Therefore, great care must be taken when examining whether the requirements for a valid consent are fulfilled.
As mentioned in Section~\ref{sec:background}, consent may need to be explicit in cases when privacy is especially at risk -- this may also require additional interactions with the user for an ``express statement of consent''.

\subsubsection{Reject}
For our classification, we use the decision type ``Reject'' as a higher level term to express  intervention 
actions for exercising the legal rights to object (Article 21 GDPR), to refuse consent or to withdraw consent (Article 7 (3)), as well as the right to opt-out of unsolicited communication of companies to their customers (Article 13 ePD).

\subsection{Level of automation}
\label{sec:levels}

Various approaches with different levels of automation have already been proposed in the literature, some of which could be considered as GDPR compliant, others as compliant but only under other legal frameworks, whilst certain approaches are likely in conflict with existing laws or ethical principles.

\subsubsection{Manual decisions}
\label{sec:manual}
The lowest level of automation is the \textbf{manual} decision that always requires a user action.

\textbf{Manual privacy permissions.}
For privacy permissions, default privacy settings that implement the most privacy-friendly options should be pre-set by the system, thus enforcing the DPbD principle (Article 25 GDPR).
Users can change them at setup or later manually on their initiative (which is however seldom the case~\cite{ausloos_guidelines_2013,watson_mapping_2015}).

Smartphone permission systems usually ask users to define permissions manually when an application is installed or first used (ask-on-install).
Ask-on-install (AOI) has the limitation that decisions are requested to be made without the user being aware of the potential contexts in which the permissions will be used.
Generally, a limitation of manual privacy permission settings is that they may need to be defined out of context and do not support privacy as contextual integrity as defined by Nissenbaum~\cite{nissenbaum2004privacy}.~\footnote{Contextual integrity can be summarized as ``a normative model, or framework, for evaluating the flow of information between agents (individuals and other entities), with a particular emphasis on explaining why certain patterns of flow provoke public outcry in the name of privacy (and why some do not).''~\cite{barth_privacy_2006}}

Note that in the council version of the upcoming ePrivacy Regulation, the possibility to express consent via browser settings may qualify as a manual permission.

\textbf{Manual privacy preferences.}
Similar to privacy permissions, default privacy preferences should implement the most privacy-friendly alternative for complying with the DPbD principle.

Privacy preferences have notably been researched in conjunction with policy languages such as PPL~\cite{ardagna2009primelife} and A-PPL \cite{azraoui2015ppl}, or Pilot~\cite{pardo_analysis_2019} for the IoT.
Manual privacy preferences are typically written or defined \textit{manually} but in a machine-readable form.

On the web, different digital signals~\footnote{Signals are defined in the literature as \textit{digital representations of how users want their personal data to be processed}~\cite{hils_privacy_2021}, but as they spawn over different categories of our classification scheme we deal with each solution separately.} have been proposed and fall under our scope of privacy preferences.
An example for signals set manually in the browser settings is ``Do Not Track'' (DNT)~\cite{kamara_not_2016}, which has however not been widely deployed due to a lack of legal mandates for its use.

\textbf{Manual consent.}
An unambiguous consent request demands a clear statement or affirmative action according to Article 4(11) GDPR, which implies that it should be given ``manually'' 
and cannot be derived implicitly.
User consent can be given at setup, or at the time and in the context when personal data is requested, or even rendered with the help of a custodian~\cite{edps_opinion_2016,european_union_agency_for_cybersecurity_data_2021,morel_enhancing_2020}.

\textbf{Manual reject.}
Reject interventions are usually done manually, e.g. opting-out may require 
users to actively change a setting in their software.
A typical example of an opt-out signal is Global Privacy Control (GPC)~\cite{noauthor_global_nodate}, which provides a way to opt-out of sharing through a browser setting (Firefox users have to toggle a setting in order to opt-out~\cite{mozilla_implementing_2021}).
GPC is enforceable under the California Consumer Privacy Act (CCPA), but cannot be considered as 
as a valid consent under the GDPR as it is an opt-out basis and not an affirmative action~\cite{human2022data}.
Manual opt-out has also been used by the Future of Privacy Forum for mobile analytics through their solution named Smart Places~\cite{future_of_privacy_forum_smart_nodate}, where data subjects had to enter their MAC address to signal their refusal to participate in data processing.

\subsubsection{Semi-automated decisions}
An intermediary level of automation for privacy decisions is their \textbf{semi-automation}.
This includes privacy decisions that are made at run-time upon dynamically created requests and/or are reacting on dynamically created recommendations. 
In this case, humans are always part of the process, and nothing is achieved without an explicit action from users.

\textbf{Semi-automated permission settings.}
Recent studies have proposed personalised and semi-automated recommendations approaches for changing users' permission settings.
These recommendations can be made by ML-based ``personalized privacy assistants'' ~\cite{liu_follow_2016,bahirat_data-driven_2018,smullen_best_2020}.
These assistants analyse users' privacy behavior and privacy personality 
in order to derive and suggest a ``privacy profile'', with privacy permission settings predicted to best suit users choices and allowing for changing to the recommended settings. 

Profiling users for their privacy preferences can, however, by itself be a privacy sensitive task. Therefore, it needs to be conducted in a privacy-preserving manner.
This can for instance be achieved if the machine-learning algorithms of personalised privacy assistants run locally under the users' control on their own devices.
Moreover, a recent federated learning approach was presented for deriving suitable privacy profiles of permission settings for users in a distributed fashion~\cite{brandao2022prediction}. This distributed architecture, which still keeps the data about the user's privacy decisions and contextual data for training the ML models locally, thus provides better privacy protection in contrast to approaches that train models based on user data centrally, where privacy preferences data must be entrusted to a central server.
Nonetheless, with a federated learning approach, locally trained models can still leak personal data, e.g. through membership inference attacks~\cite{shokri2017membership}. Hence, further privacy-preserving measure (e.g. differential privacy) need to be implemented in addition.  

Personalised privacy assistants also often involve privacy nudges~\footnote{Nudges are ``any aspect of the choice architecture that alters people’s behavior in a predictable way without forbidding any options or significantly changing their economic incentives''~\cite{thaler2009nudge}, this concept gained a significant traction when applied to make better privacy decisions~\cite{acquisti_nudges_2017}.} designed to motivate users to revisit their earlier decisions~\cite{liu_follow_2016}.
Indeed, prior research has emphasized the role that risk awareness has when building systems aiming to aid peoples’ privacy decisions~\cite{ferreyra_persuasion_2020,samat2017format,distler_systematic_2021}.
These nudges however raises ethical questions, e.g., in regard to the undermining of users' autonomy, as discussed in~\cite{veretilnykova2021nudging}, although researchers have introduced ethical guidelines for the design of privacy and security nudges~\cite{renaud_ethical_2018}.

Access control rules/permission settings can also be created dynamically ``on the fly'' in the form of sticky policies.
For instance, with the PrimeLife policy Language PPL, the matches of preferences with a data controller’s policy may result (with the user's consent) in a ``sticky policy'', which defines access control rules that oblige the data controller (and its data processors)~\cite{ardagna2009primelife}.

Mobile app permissions may also be defined dynamically, when an app is first attempting to access a “dangerous” permission type such as location or contacts (ask-on-first-use).
However, while the current context for which the permissions are requested is apparent with ask-on-first-use (AOFU), the context may still change in future situations when the permissions will be used~\cite{wijesekera2017feasibility}; the limitation previously addressed for AOI hence still applies.

\begin{figure}[t]
	\centering
	\includegraphics[width=8cm]{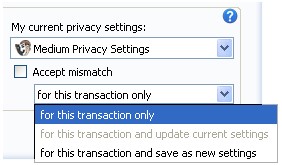}
	\caption{User interface elements for ``On the fly" privacy preference management \cite{angulo2012towards}}
	\label{fig:on-the-fly}
\end{figure}

\textbf{Semi-automated preference settings.}
Managing privacy preferences could be done dynamically ``on the fly'': if users make a decision differing from their defined preferences, they could be asked whether they would like to update their preferences accordingly.
For instance, Angulo et al.~\cite{angulo2012towards} suggested a usable \textit{“on-the-fly” policy management} for PPL.
In case of a mismatch between the user's preferences and the controller's policy, the user is asked in a consent form if they would anyhow like to accept this mismatch and disclose the data for this transaction only, or if they want to accept such a mismatch for all future transactions as well. 
In the latter case, it is also suggested that users’ preferences are updated accordingly (see Figure~\ref{fig:on-the-fly}).
Similarly, updates are also suggested if the user does not consent to data disclosures even if there is no mismatch.

Das et al. envisioned in~\cite{das_personalized_2018} a privacy assistant for the IoT managing privacy preferences in a semi-automated way.
Such assistant would be able to model preferences in a way similar to the mobile assistant proposed in~\cite{liu_follow_2016}.
It could then detect mismatches between users' preferences and privacy policies of IoT resources, and warn them in such cases.

\textbf{Semi-automated and dynamic consent.}
\label{sec:dynamic}
A \textit{dynamic consent} can be defined as a regular consent that  will be \textit{requested} in a specific context any time after an initial consent was collected, particularly for authorising incremental changes to the previously given consent, e.g. in case that the data controller would also like to process the data for other purposes or to change its policy~\cite{schlehahn2020}. 
Moreover, if it is detected that, in the current context, the data to be collected/processed classifies as special categories of data (for instance due to inference~\cite{malkin_contextual_2023}), an explicit consent from the user should be obtained dynamically.

\begin{figure}[t]
	\centering
	\includegraphics[width=8cm]{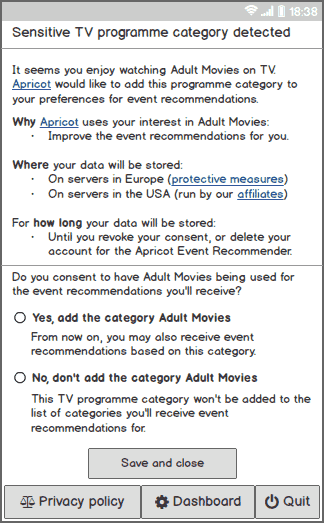}
	\caption{Scenario for requesting dynamic consent for ``re-purposing'' data: A user who has previously consented to profiling of their TV viewing behaviour and for using it for TV recommendations is now  asked for giving their dynamic consent for allowing to add and process a sensitive TV profile category (interest in adult movies that may reveal the user's sexual preferences) for receiving event recommendations \cite{schlehahn2020}.}
	\label{fig:dynamic-consent}
\end{figure}

Figure~\ref{fig:dynamic-consent} shows a user interface prototype for requesting dynamic explicit consent for using sensitive data categories in a commercial use case scenario developed within the SPECIAL and Privacy\&Us EU projects. 

Dynamic consent can be seen as a special form of \textit{Just-in-time click-through agreements} (JITCTAs), which ere initially presented by Patrick and Kenny~\cite{patrick2003privacy}. JITCTAs provide a short contextualised (``just-in-time'') privacy notice for obtaining consent through a concise dialogue specific to a certain data practice. They are triggered when informed consent becomes relevant for the user, e.g. in case that data practices are considered sensitive or unexpected~\cite{kobsa2005contextualized, schaub2015design}.

While it is not a different legal concept than traditional consent, this approach coming from the biomedical domain~\cite{prictor_dynamic_2020} offers new interesting perspectives for personal data management.
For instance, Asghar and Russello~\cite{asghar2012flexible} proposed a high-level architecture to request consent dynamically through a process engine and a consent evaluator, depending on previously given consents.
Under a different name, a protocol for the negotiation and the communication of consent was proposed for the IoT~\cite{morel_enhancing_2020}.
In the suggested solution, if a previously manually defined \textit{data subject policy} does not match the controller's privacy policy, data subjects are then notified of this mismatch, and can change their data subject policy to communicate consent.
If data subjects choose not to change their policy, they can start a negotiation to lower the expectations of the controller.
If the negotiation succeeds, data subjects can consent to a less demanding controller's privacy policy.

\textbf{Semi-automated reject.}
A partial automation of opt-out mechanisms did not seem to foster an important body of research, although it seems that some websites are proposing a form of simplification of opt-out choices~\cite{habib_empirical_2019}.
However, we stress that Article 21 (5) GDPR specifically mentions, in concrete, that the right to object may be exercised ``by automated means using technical specifications''.

\subsubsection{Fully automated decision}
\label{sec:fully}
Finally, a privacy decision can be also fully automated.
In that case, the automated decision feature, especially if ML-based, should first be actively enabled by users, also because they have the right, under the GDPR, not to be subject to a decision based solely on automated processing (Article 22).

\textbf{Automated privacy permission settings.}
Permission settings or access control rules can in principle be automatically overwritten, e.g., by an intrusion prevention system, which revokes access control rules for protecting users if privacy intrusions or high risks are detected.

However, automatically granting permission/access rights in access control rules without the users' involvement may be problematic, as it may enable access to the users' data without their explicit permission or consent.

Nevertheless, Wijesekera et al.~\cite{wijesekera2017feasibility} have developed a ML-based approach, accurately inferring privacy ``preferences'' (which are rather permissions in our classification) based on the user’s past decisions and behavior, in order to automatically grant appropriate smart phone permission requests.
Their approach also considers context changes without user intervention, it denies inappropriate requests, and only prompt users when the system is not certain of their preferences.

\textbf{Automated privacy preference settings.}
Privacy preferences could also be automatically adapted based on the users’ behavior and detected personality, or to protect users in risky situations.
Automating the settings of privacy preferences that are used to increase transparency is not conflicting with the GDPR's consent requirements, as preferences only express ``wishes'' and users are still required to consent manually.
Still, it is questionable whether it should be possible to also automatically adapt settings to more generous ones without involving users, as this approach could be seen as in conflict with the DPbD principle (Article 25 GDPR).

\textbf{Automated consent.}
While automation of consent arguably facilitates decision making and can significantly reduce user burden, the lawfulness of a fully automated approach applied to consent in an EU context is highly questionable, as an automated consent without user interaction is neither informed nor freely given (see Section~\ref{sec:background}).

Despite legal concerns regarding GDPR compliance, research on automated consent models predicting users' data-sharing decisions have been conducted, mostly by non-European researchers.
For instance, Colnago et al.~\cite{colnago_informing_2020} conducted interviews, in which participants suggested that privacy assistants could be able to recognize when users face similar consent decisions in order to implement the decision again.
Colnago et al. have shown that these automated models are negatively perceived by users that desire to stay in control, and they therefore also suggest an automated consent that ``holds data in escrow'', so that it is not immediately available to the requesting party, given the user time to review and object. 
However, this proposed opt-out option would not constitute a valid consent pursuant to the GDPR.

Also Mendes et al.~\cite{mendes2022enhancing} have recently researched automated privacy decision systems based on ML-derived personalised predictions of privacy
decisions for Android systems, but conclude as well that such automated decisions responding to a permission request might not constitute
legal consent pursuant to the GDPR.  Therefore, they rather suggest using semi-automated approaches for recommending predicted privacy decisions to users.

On the web, several cookie consent tools are also automating consent decisions.
For instance, the extension ``I don't care about cookies''~\cite{kladnik_i_nodate} removes cookie banners (``cookie warnings''), and it is programmed to automatically accept the ``cookie policy'', that is, to consent.
Firefox Cookie Banner Handling~\cite{firefox_firefox_2022} similarly dismisses banners by refusing consent whenever possible, but will consent on behalf of users if no other choice is apparent.
In this last case, there is no valid consent pursuant to the GDPR, as this is not an affirmative action.~\footnote{It appears that Firefox now supports two version of cookie banner handling: \textit{reject all} or \textit{reject all or fall back to accept all}, see~\cite{mozilla_firefox_cookie_2023}. Note that the first option does not convey consent.}

Interestingly, as Santos et al. \cite{Santos23} point out and discuss, the ePrivacy Regulation proposals (in the versions of the Commission, Parliament and Council) refer to the possibility for consent to exist through technical software-settings, though its current progress does not yet define  such a software-settings based consent decision as legally binding. 

\textbf{Automated reject}
In contrast with tools for automating consent that are not GDPR-compliant, automated tools to refuse or withdraw consent for protecting the users -- which could be ML-supported --, could be legally compliant solutions.
For instance \textit{consent revocation} could be proposed dynamically if a privacy risk for compromising the users' data is detected automatically, or if it is detected that the users' automatically inferred privacy personality is not matching their previous consent decisions.

Consent-O-Matic~\cite{nouwens_dark_2020} is an example of a cookie consent tool
which only dismisses cookies and leaves the notice if it cannot automatically deal with the banner, so that the choice is up to the user.

 Santos et al. \cite{Santos23} also discuss the possibilities of automating the withdrawal of consent or automated means to exercise the right to object via privacy signals. The Advanced Data Protection Control (ADPC)~\cite{human2021advanced} supports both the communication and automation of exercising consent withdrawals and objectives, while the IAB Transparency and Consent Framework (TCF)~\cite{iab_europe_tcf_nodate} supports the communication of objections. 

\section{Concluding remarks}
\label{sec:discussions}

Automation, or semi-automation for the most part, has the potential to increase user control by enhancing usable transparency
and enable fine-grained and contextual controls, e.g. via dynamic consent.
Recent research has demonstrated that ML techniques can accurately predict the users' privacy choices with more than 95\% (see~\cite{wijesekera2017feasibility}) and may lead to decisions better matching the users' privacy interests than manually made decisions.
This aspect is to be considered in conjunction with the fact that users are often practically not well informed, because it is simply too time-consuming and demanding for them in practice to carefully study privacy notices as part of consent forms.~\footnote{See also~\cite{mcdonald2008cost} that estimated that the time required for reading policies in a year would on average exceed 200 hours for a user.}
A cognitive overload due to frequently appearing requests for manually made decisions  contributes to this problem. 
Also since privacy is usually only a secondary goal for users~\cite{whitten1999johnny}, they hastily happen to make choices resulting out of habituation or are falling for dark patterns.~\footnote{Dark patterns are ``instances where designers use their knowledge of human behavior (e.g., psychology) and the desires of end users to implement deceptive functionality that is not in the user’s best interest''~\cite{gray_dark_2018}.} 

Moreover, especially for IoT devices with restricted access to user interfaces that are needed for making choices~\cite{castelluccia_enhancing_2018,morel_enhancing_2020}, automation could be a means for protecting users' privacy more effectively.

On the other side, the users' autonomy and control need to be protected for complying with GDPR requirements, ethical principles, and for enabling trust by users that would like to retain agency.
A fully automated approach is in conflict with GDPR requirements on consent, and could defeat the purpose of enhancing transparency for privacy preferences and permissions settings.
Moreover, automatically inferred privacy preferences have to be carefully crafted, they are a double-edged sword which could provide tailored settings as well as diminish data subjects' autonomy~\cite{vaassen_ai_2022}, and can conflict with the data protection by default principle.
Only the automation of reject decisions appears to meet lawfulness requirements and may enhance usability at the same time. 

Nonetheless, since users' privacy decisions (including ``reject'' decisions) can never be perfectly predicted with 100\% precision, fully automated decisions may not always meet the users' expectations and can therefore only enhance usability if they are very transparent and can also be easily corrected by users (which does not palliate their eventual failure to comply with legal requirements).


In a nutshell:
\begin{itemize}
    \item Manual decisions set the onus on users -- though they stay formally in control, they may practically not be capable to make well-informed decisions reflecting their privacy interests for a huge amount of decision requests that users are typically confronted with;
    \item Fully automated decisions are almost always in conflict with legal requirements (with the exception of reject decisions);
    \item Semi-automated decisions can meet both ends of usability and legal compliance, but great care must be taken to preserve agency of users when devising solutions of that trend.
\end{itemize}

We specifically raised in Section~\ref{sec:intro} the question of consent, whose management is currently being subject to technical changes in both the web and the IoT.
Cookie consent tools and signals participate in the automation of privacy decisions on the web, while privacy assistants tackle this question in the IoT.

In our future project work, we plan to develop usable permission systems for another type of environment, that is, IoT Trigger Action Platforms (TAPs).
IoT TAPs are an emerging technology that gained traction in academia over the recent years, but little research has been conducted on enhancing privacy in this context.
IoT TAPs however possess specific features impacting privacy: these environments combine the restriction of IoT devices (e.g., limited interfaces) with a connection to data-hungry web services.  
Our envisioned system would combine a privacy permission ``on the fly'' approach with semi-automatically requested dynamic consent decisions. 
This enables context-specific and fine-grained choices that can be requested especially in the context when data becomes sensitive. If for instance the users' location can in a certain context (e.g. time when a religious service or political event takes place at a certain location) become sensitive information, the users' dynamic consent for revealing their location data in that context could be requested. If users deny explicit consent, they can be asked if they would like to adapt their permission settings accordingly. Alternatively, recommendations for consent decisions or adaption of permissions settings can be given in that context that are corresponding to the privacy profile of a specific user.
Our approach should support data subjects in their privacy decisions while meeting legal requirements of the GDPR. 

The classification scheme provided in this paper can be used as a means to determine the legal compliance of existing and future applications, in a way that would solve the tension between 1) the reduction of the cognitive burden on data subjects while 2) respecting their best interests and rights and 3) ensuring principles such as data protection by design and default.

Future work may include an exhaustive compilation of the solutions addressing automation of privacy decisions, as well as the investigation of the impact on automation on other types of rights (such as the right to portability, to access data, etc).

\section*{Acknowledgments}
Thanks are due to Cristiana Santos for her valuable feedback on the legal aspects of the paper. 
This work was partially supported by the Wallenberg AI, Autonomous Systems and Software Program (WASP) funded by the Knut and Alice Wallenberg Foundation.

\bibliographystyle{plainurl}
\bibliography{mybibliography}

\end{document}